\documentclass[fleqn,twoside]{article}
\usepackage{espcrc2,epsfig}

\newcommand{\ba}{\begin{eqnarray}}
\newcommand{\ea}{\end{eqnarray}}
\newcommand{\fig}{Fig.~}

\newcommand{\eq}{Eq.~}
\newcommand{\nr}[1]{(\ref{#1})}
\newcommand{\nn}{\nonumber \\}
\newcommand{\fr}[2]{{\frac{#1}{#2}\,}}
\newcommand{\msbar}{{\overline{\mbox{\rm MS}}}}
\newcommand{\lambdamsbar}{{\Lambda_{\overline{\rm MS}}}}
\renewcommand{\(}{\left(}
\renewcommand{\)}{\right)}
\newcommand{\lb}{\left\{}
\newcommand{\rb}{\right\}}
\newcommand{\lk}{\left[}
\newcommand{\rk}{\right]}
\newcommand{\ld}{\left.}
\newcommand{\rd}{\right.}
\renewcommand{\d}{\delta}

\renewcommand{\l}{\lambda}
\newcommand{\6}{\partial}
\newcommand{\cF}{{\cal F}}

\newcommand{\rmi}[1]{{\mbox{\scriptsize #1}}}
\newcommand{\cFms}{{\cal F}_{\rmi{$\msbar$}}}
\newcommand{\bG}{{\beta_{\rm G}}}

\newcommand{\LAv}[1]{\left\langle #1 \right\rangle_{\rm L}}
\newcommand{\tr}{{\rm Tr}}

\title{Measuring infrared contributions to the QCD pressure\thanks{
Work supported in part by the EU TMR network 
ERBFMRX-CT97-0122.}}
\author{K. Kajantie\address[Hki]{Department of Physics,
P.O.Box 64, FIN-00014 University of Helsinki, Finland}, 
M. Laine\address{Theory Division, CERN, CH-1211 Geneva 23,
Switzerland}, 
K. Rummukainen\address{NORDITA, Blegdamsvej 17,
DK-2100 Copenhagen \O, Denmark},
Y. Schr\"oder\addressmark[Hki]\thanks{Present address: 
Center for Theoretical Physics, MIT, Cambridge, MA 02139, 
USA.}\thanks{talk given at {\em Lattice 2001}, Berlin, August 19-24, 2001.}}

\begin{document}

\begin{abstract}
For the pressure (or free energy) of QCD, four-dimensional (4d)
lattice data is available at zero baryon density up to a few times 
the critical temperature $T_c$. Perturbation theory, on the other hand,
has serious convergence problems even at very high temperatures. 
In a combined analytical and three-dimensional (3d) lattice method, 
we show that it is possible to compute the QCD pressure from 
about $2 T_c$ to infinity. The numerical accuracy is good enough
to resolve in principle, e.g., logarithmic contributions related to 
4-loop perturbation theory.
\end{abstract}

\maketitle


\setcounter{footnote}{0}

\noindent{\bf Introduction.}
The properties of QCD matter are expected to change above 
a critical temperature $T_c\sim $ 200 MeV. 
While the low-temperature phase is governed by bound states,
such as mesons, the high-temperature phase should, due to
asymptotic freedom, look more like a gas of free quarks and gluons.
Any observable witnessing this change is therefore a potential 
candidate for direct or indirect
measurements in heavy-ion collision experiments.
One such observable is the free energy density, or pressure 
of the system, since according to the Stefan-Boltzmann law 
(i.e.~neglecting interactions), 
it is proportional to the number of effective degrees of freedom. 

For vanishing baryon density, at a temperature $T$ and volume $V$,
the free energy density $f$ is simply given by the functional integral
\ba
f = -\fr{T}V\ln\!\int{\cal D}[A\bar\psi\psi]
e^{-\!\int_0^{1/T} \!\!\!\!\!d\tau \int_V \!\!d^3x \, 
{\cal L}_E[A\bar\psi\psi]} \;,
\label{int}
\ea
where ${\cal L}_E$ is the standard QCD Lagrangian.
For $V\to\infty$, the pressure $p$ is given by $p=-f$. 

The most direct way to evaluate \eq\nr{int} is
to measure it numerically on the lattice. This
has been done by a number of groups, for zero as well as
non-zero number of fermion flavours $N_f$ (e.g.~\cite{lat,latNf}). 
The general picture emerging is the following: The pressure 
rises sharply in the interval $(1-2)T_c$, to level off at a few times
$T_c$. At the highest temperatures used in the simulations,
typically $(4-5)T_c$,
the deviation from the Stefan-Boltzmann limit is about $15\%$.
At even higher temperatures, the pressure is then expected to
asymptotically approach the ideal-gas value
$p_0(T)=(\pi^2 T^4/45)(N_c^2\!-\!1+(7/4)N_cN_f)$, where $N_c$ denotes 
the number of colours. 

The $15\%$ deviation cannot be systematically understood 
in terms of (finite-temperature) perturbation theory. 
While the expansion is known analytically to 5th order 
in the gauge coupling $g$~\cite{zk}, convergence properties 
are extremely poor.
Therefore, in the past
few years a lot of effort has gone into refined and/or alternative 
analytic approaches \cite{bir}. 
A general 
feature of these works is the suppression of infrared
effects. While this suppression does not seem to be crucial in the 
computation of the pressure, one might not be satisfied with accepting it as
an {\em ad hoc} assumption.
The aim of this talk is to briefly review our 
framework of resumming the infrared contributions to the 
pressure to all orders \cite{fqcd}. 


\mbox{}

\noindent{\bf Pressure via effective theory.}
A way to understand the poor convergence of the ordinary perturbative
expansion is the observation that when $gT\ll\pi T$,
the system undergoes dimensional reduction (see, e.g.,~\cite{bn} and
references therein). 
In the case of QCD, 
the effective theory is a
3d SU($N_c$) + adjoint Higgs model:
\ba \label{adjH}
{\cal L}_{\rm 3d} = 
\fr14 F_{ij}^2 +\!\fr12\! [D_i,A_0]^2 +\!\fr12\! m^2_3 A_0^2 
+\!\fr14\! \lambda_3 A_0^4. 
\ea
The parameters ($g_3$, $m^2_3$, $\l_3$) are related to the physical 
parameters of the full 4d theory ($T$, $\lambdamsbar$)
by perturbatively integrating out the hard modes ($\sim\pi T$).
Below, we will work with the dimensionless variables 
$x=({\l_3}/{g_3^2})$ and $y=({m^2_3}/{g_3^4})$, 
known at next-to-leading order~\cite{ad}.

The effective theory is confining,
hence non-perturbative \cite{nonpert}. 
Therefore, the only way to systematically include the infrared 
contributions to the pressure is to treat ${\cal L}_{\rm 3d}$ on the
lattice.

Let us rewrite the pressure, up to hard-scale $g^6$ contributions, as
\ba \label{newNot}
\fr{p(T)}{p_0(T)} = 1 -\fr{5x}2 -\fr{45}{8\pi^2} \(\fr{g_3^2}T\)^3 
\lk {\cal F}_{\overline{\rm MS}}(x,y) 
\vphantom{\fr11}\rd\nn\ld{}
-\fr{24y}{(4\pi)^2} 
\(\ln\fr{\bar\mu_{\rm {3d}}}T +\d\) \rk \;,
\ea
where $\d\sim10^{-4}$, 
$p_0(T)$ is the Stefan-Boltzmann value, 
and the dependence on the scale $\bar{\mu}_{\rm {3d}}$, 
which originates from an infrared divergence of the 4d part, cancels
against an ultraviolet term\footnote{This is precisely 
the way the effective theory is set up: dependence on a matching scale
has to cancel.} 
in the dimensionless 3d free energy density 
${\cal F}_{\overline{\rm MS}}(x,y)$. 
We write ${\cal F}_{\overline{\rm MS}}(x,y)$ as
\ba
{\cal F}_{\overline{\rm MS}}(x_0,y_0) 
+\!\int_{y_0}^{y} \!\! dy \( 
\fr{\partial {\cal F}_{\overline{\rm MS}}}{\partial y}\! + \fr{d x}{dy} 
\fr{\partial {\cal F}_{\overline{\rm MS}}}{\partial x}
\) \;,\label{3dF}
\ea
and aim at measuring it on the lattice. 
This requires a determination of
the quadratic and quartic Higgs field condensates 
(which equal the partial derivatives under the integral), as
well as a (4-loop) perturbative computation in lattice regularization, 
to match to the $\msbar$ scheme. We choose to fix the integration
constant ${\cal F}_{\overline{\rm MS}}(x_0,y_0)$
at high temperatures $T\sim 10^{11} T_c$, 
where one is confident that continuum
perturbation theory converges\footnote{
$T_c$ is of the order of $\lambdamsbar$, with a coefficient 
measurable on 4d lattices.}.

\begin{figure}[t]

\epsfig{file=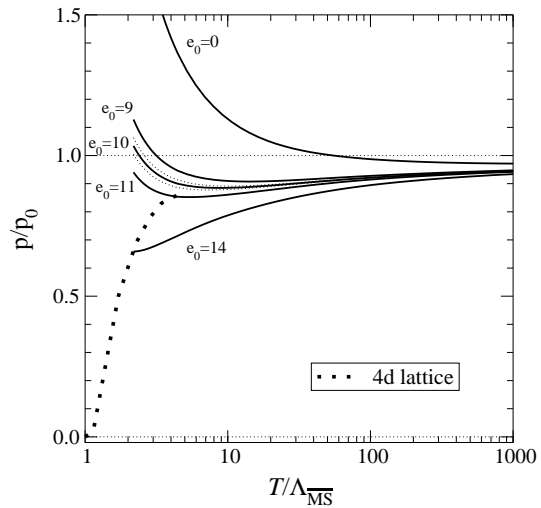,width=70mm}

\vskip -0.8truecm

\caption{The pressure after the inclusion of the infrared 
part according to \eq\nr{3dF} (from~Ref.~\cite{fqcd}). 
Statistical errors are shown only for $e_0=10$.}

\vskip -0.5truecm

\label{fig:press}
\end{figure}

As a first result obtained along the strategy outlined,
\fig\ref{fig:press} shows the normalized pressure. The integration constant
has been fixed perturbatively on the 3-loop level, allowing for 
an additional constant $e_0$, which represents an (up to now) 
unknown $g_3^6$ contribution. In principle, this constant can be
determined in a setup equivalent to the above, after splitting
off its perturbative part: A further reduction
relates $e_0$ to the free energy of 3d pure gauge theory
(see, e.g.,~\cite{bn}), which can 
be determined on the lattice.
On the lattice side, we have only included the 
quadratic scalar condensate in \fig\ref{fig:press}, 
due to reasons explained in the next section; however,
at $T\to T_c$ the quartic one will become important as well. 

\mbox{}

\noindent{\bf 3d lattice measurements.}
Let us now discuss in more detail how the actual measurement of the 
3d condensates is carried out.
We need to relate lattice and $\msbar$ regularization schemes.
Here, the super-renor\-ma\-liza\-bility 
of the 3d theory plays an important role:
there are only a finite number of divergent terms, such that analytic relations
can be computed exactly near the continuum limit~\cite{framework}. 
Writing
$\bG=({2N_c}/{g_3^2 a})$ where $a$ is the lattice spacing, the observables
we need behave schematically like
\ba
\6_y\cFms \sim\! \lim_{\bG\to\infty}\!\lb \LAv{\tr A_0^2}\! +\!\bG 
+\!\ln\bG +\!1 \rb,
\label{A2cond}\\
\6_x\cFms \sim \lim_{\bG\to\infty}\lb \LAv{(\tr A_0^2)^2} +
\bG^2 \mbox{\hspace{36pt}}
\rd\nn\ld\vphantom{1_1^1}{}
+\bG\ln\bG +\bG +\ln\bG +1 \rb. \label{A4cond}
\ea
For \eq\nr{A2cond} all three coefficients are known~\cite{lr}. 
In the case of the quartic condensate, \eq\nr{A4cond}, all we know
so far are the four divergent terms: the weakest logarithmic 
divergence can be obtained in a 4-loop continuum calculation,
but the 4-loop lattice constant is still unknown. This is precisely the
reason why in \fig\ref{fig:press} only the effect of $\LAv{\tr A_0^2}$
is included. While the graphs needed can be systematically 
generated~\cite{sd}, carrying out the lattice integrals remains
a major challenge.

\begin{figure}[t]

\epsfig{file=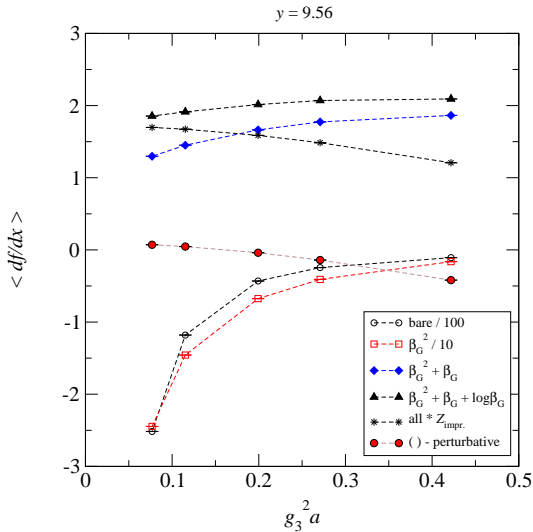,width=70mm}

\vskip -0.8truecm

\caption{A measurement of one of the partial derivatives
needed in \eq\nr{3dF}, $\6_x\cF$, at a fixed~$y$,
as a function of the lattice spacing $g_3^2 a$.
Note that the raw data (open circles) is scaled
by a factor 1/100. 
The other curves result from subtracting various divergences, 
\eq\nr{A4cond}; taking into account improvement \cite{Guy}; and finally
(filled circles), subtracting the perturbative value.} 

\vskip -0.5truecm

\label{fig:dfdx}
\end{figure}

Nevertheless, we can show how the data involving the quartic condensate
behaves. In \fig\ref{fig:dfdx}, we demonstrate for one specific temperature 
that the series in \eq\nr{A4cond}
indeed regulates the lattice operator as $g_3^2 a\to 0$. Indeed, 
taking into account successive terms of the series, one observes
how the divergence at small $a$ becomes weaker. 
Note that while the continuum
limit can be taken already now, the actual value of $\6_x\cFms$ will have
to await the above-mentioned 4-loop computation of the finite
term in lattice regularization.


\mbox{}

\noindent{\bf Conclusions.}
We wish to point out two trends seen in 
\fig\ref{fig:press}: 
First, the outcome is sensitive to the non-perturbative 
parameter $e_0$, which in principle can be determined by additional
computations. Clearly, there exists a range for that parameter 
which leads to a sensible result. 

Second, comparing with \fig{1} of \cite{fqcd}, 
at $T>30\lambdamsbar$ the curves for $g^5$ (denoted by $y^{\fr12}$ there) 
and $e_0=10$ fall
almost on top of each other, signalling a cancellation of all higher-order 
terms (determined by the quadratic Higgs condensate) against the large 
non-perturbative $g^6$ contribution. Hence, in this temperature range the
pressure is indeed dominated by ultraviolet effects.

Let us also remark that 
as demonstrated in \fig\ref{fig:dfdx}, 3d lattice
results involve always a systematic extrapolation to the continuum limit, 
and are precise enough to resolve, e.g., logarithmic effects related
to 4-loop perturbation theory. 

We end by noting that the 
inclusion of $N_f$ fermion flavours as well as a baryon chemical
potential $\mu_b$ pose no further complications, and hence provide 
for a natural extension of this investigation.


\end{document}